\documentclass[10pt,conference]{IEEEtran}
\IEEEoverridecommandlockouts
\usepackage{cite}
\usepackage{hyperref}
\usepackage{mathtools}
\usepackage{amsmath,amssymb,amsfonts}
\usepackage{graphicx}
\usepackage{textcomp}
\usepackage{soul}
\usepackage[table]{xcolor}
\usepackage{etc}
\usepackage{algorithm}
\usepackage{algpseudocode}
\usepackage{booktabs}
\usepackage[normalem]{ulem}
\usepackage{subcaption}
\usepackage{enumitem}
\usepackage{balance}

\usepackage{multirow,array}

\newcommand{\llmc}{\textit{CLM}\xspace}
\newcolumntype{C}[1]{>{\centering\arraybackslash}p{#1}}

\definecolor{darkgreen}{rgb}{0.0, 0.5, 0.0}

\newcommand{\fonescore}{\emph{F1-score}\xspace}
\newcommand{\cic}{\textbf{\textit{CIRC}}}
\newcommand{\ote}{\textbf{\textit{OTER}}}

\newcommand{\framework}{{\sc brace}}

\def\BibTeX{{\rm B\kern-.05em{\sc i\kern-.025em b}\kern-.08em
    T\kern-.1667em\lower.7ex\hbox{E}\kern-.125emX}}

\definecolor{codegreen}{rgb}{0.1,0.7,0.1}
\definecolor{codegray}{rgb}{0.5,0.5,0.5}

\newtcolorbox{rqbox}[1][]{
    colback=green!10,
    colframe=codegreen,
    arc=1mm,
    boxrule=0.5pt,
    coltitle=black,
    fonttitle=\bfseries,
    title=#1
}

\newtcolorbox{rqboxgray}[1][]{
    colback=codegray!10,
    colframe=codegray,
    arc=1mm,
    boxrule=0.5pt,
    coltitle=black,
    fonttitle=\bfseries,
    title=#1
}

\definecolor{green}{rgb}{0.0, 0.5, 0.0}

\begin{document}

\title{Smart but Costly? Benchmarking LLMs on Functional Accuracy and Energy Efficiency}

\author{
\IEEEauthorblockN{Mohammadjavad Mehditabar}
\IEEEauthorblockA{\textit{Dalhousie University}\\
Halifax, Canada\\
javad@dal.ca}\\
\IEEEauthorblockN{Antonio Mastropaolo}
\IEEEauthorblockA{\textit{William \& Mary} \\
Williamsburg, Virginia, USA
 \\
amastropaolo@wm.edu}
\and
\IEEEauthorblockN{Saurabhsingh Rajput}
\IEEEauthorblockA{\textit{Dalhousie University}\\
Halifax, Canada\\
saurabh@dal.ca}\\
\IEEEauthorblockN{Tushar Sharma}
\IEEEauthorblockA{\textit{Dalhousie University}\\
Halifax, Canada\\
tushar@dal.ca}\\
}

\maketitle

\begin{abstract}

The rapid advancement of AI technologies and their accelerated adoption in software development necessitates a systematic evaluation of their environmental impact alongside functional correctness. 
While prior studies have examined sustainability in large language models, existing approaches lack systematic frameworks for evaluating accuracy-energy trade-offs in Code Language Models (\llmc{}s).
In this paper, we present a framework, \framework{}, to benchmark \llmc{}s on a unified scale of energy efficiency and functional correctness (referred to as accuracy).
We benchmark 22 state-of-the-art models on code generation and summarization tasks, proposing two rating methods: Concentric Incremental Rating Circles (CIRC) and Observation to Expectation Rating (OTER). 
CIRC provides deterministic Euclidean-based rankings with static trade-offs that are robust to outliers, and OTER offers trend-aware evaluation with dynamic trade-offs that capture the complex correlation between energy and accuracy, each offering a distinct perspective and addressing the problem in a unique way. These rating methods enable us to rate LLMs on a 1-5 scale reflecting their combined capabilities in terms of energy efficiency and functional correctness. 
Our analysis reveals models generally perform better in the code summarization tasks as they are not enforced to generate a grammar-based and syntactically correct output. Also, we find that models' size does not have a significant impact on their ratings, indicating that if models utilize their parameters efficiently, they can be ranked higher on these scales. 
The proposed \framework{} framework empowers practitioners to make evidence-based model selections that balance sustainability with task requirements, guiding rating choice---CIRC for deterministic comparisons or OTER for trend-aware evaluation---based on deployment priorities.

\end{abstract}

\begin{IEEEkeywords}
Code Language Models, benchmarking, sustainability rating, energy consumption, green AI.
\end{IEEEkeywords}

\section{Introduction}
\label{intro}

Generative Artificial Intelligence (GenAI) tools and platforms,
empowered by Large Language Models (LLMs), have seen widespread adoption in the public domain~\cite{bright2024} thanks
to their ability to generate human-like text, process complex queries, and assist users performing various tasks~\cite{yao2023react}.
In the context of software engineering (SE),
Large Language Models for Code (\textit{CLMs}) 
such as Codex~\cite{chen2021evaluating}, the model that powers GitHub Copilot~\cite{github_copilot}, are trained on vast corpora of source code and natural language data to facilitate and streamline software development activities.
These models can generate, analyze, debug, and optimize code across multiple programming languages~\cite{Hou2024,CLM_strength,Fan2023}, fundamentally transforming how developers approach programming tasks.
As a result, developers are increasingly integrating such solutions into their day-to-day programming workflows~\cite{ZUBAIR2025,Qiu2025,mastropaolo2025triumph}, with adoption rates continuing to rise across diverse development environments and team structures.

\llmc{}s are evolving rapidly, increasingly closing the task-specific capabilities gaps between models and humans while expanding their reach to new software engineering tasks~\cite{Wang2025,Hou2024}.
This evolution is driven primarily by scaling laws through increased model size that enabled \llmc to better capture the latent semantic space of software engineering tasks, and, consequently, improved task accuracy.
However, this development direction carries significant environmental costs that are often overlooked. Scaling models' parameter size increases considerable energy consumption~\cite{brownlee2021exploring,strubell2020energy,nguyen2024towards},
and carbon emission~\cite{li2025making,faiz2023llmcarbon}. 
Mitigating this environmental impact has become a critical concern, prompting research on sustainable and efficient solutions that aim at optimizing energy consumption while ideally preserving the benefits of evolving AI models\cite{Georgio2022,BolonCanedo2024,afrin2025quantization}.

A key direction within this effort is benchmarking LLMs on sustainability-related metrics.
For example, the AI energy score~\cite{aienergyscore-leaderboard} benchmarks LLMs based on their energy consumption on a specific set of tasks and assigns a star-rating from $1$ to $5$.
However, benchmarking LLMs only based on a sustainability metric, such as energy consumption, presents an incomplete perspective that neglects the fundamental role of functional performance, which is key for practical adoption.
A highly energy-efficient model that achieves only mediocre accuracy levels can be a ``show-stopper'' to users who prioritize task completion over resource efficiency. Conversely, an accurate but energy-intensive model may prove unsustainable at scale. Though a few studies~\cite{husom2025sustainable,alizadeh2024analyzing} have applied Pareto optimality to identify non-dominated points --- where no model outperforms another in both energy and accuracy --- this approach suffers from three major limitations.
First, it highlights only the dominant points, ignoring the rest. Second, without a mathematical function to combine the objectives, comparing points is unfeasible. Lastly, points on the frontier can differ significantly in overall performance, as models with extremely high value in one objective are not rewarded or penalized fairly.
Therefore, combining functional performance (\ie{} task accuracy) and efficiency (\ie{} energy consumption) is necessary for holistic model evaluation that reflects real-world deployment constraints.

There has been some attempts to combine accuracy and energy aspects for benchmarking LLMs.
In this direction, Kaplan \etal{}'s ~\cite{kaplan2025} attempts to jointly incorporate \fonescore and carbon emissions by computing the ratio between normalized carbon emission and F1-score. However, ratio metric assumes a linear relationship between two objectives which overlooks the non-linear and complex interaction between two metrics. Similarly, Jegham \etal{}~\cite{jegham2025hungry} propose a scoring scheme built on top of DEA~\cite{CHARNES1978429} that  measures how effectively each model converts sustainability-related factors (\ie{} inputs of the technique) into multiple outputs represented by varying accuracy levels. Despite its novel conceptual framework, the approach suffers from overfitting~\cite{guillen2024improving}, sensitivity to outliers~\cite{cazals2002nonparametric}, and other ranking issues~\cite{tone2001slacks}. Lastly, though mean-based variants, \textit{e.g.,} arithmetic mean and harmonic mean, were not employed in energy-accuracy context, they are widely adopted in different domains to combine objectives~\cite{vandierendonck2017comparison,schafer2020teaser,Sauro2005}. However, similar to ratio metric, their scoring mechanism operates independently of objective correlation and may unfairly favor cases with extremely low performance in one aspect, or conversely, over-penalize low values in the other---thereby failing to establish a robust method that jointly evaluates the energy consumption and accuracy of models---an aspect that remains unexplored in the literature.

To fill this research gap, we present \textbf{\framework{}}, the first unified benchmark for \llmc{}s that integrates the key dimensions underlying the evaluation of software engineering tasks---specifically \textit{functional correctness}, modeled through accuracy, and energy consumption, representing the \textit{non-functional} dimension of model performance.
We develop a benchmarking framework that seamlessly measures the energy and accuracy of specified models. Given the dual-objective optimization nature of the problem, we model it as a Multi-Criteria Decision Making (MCDM)~\cite{zionts1979mcdm} problem and introduce two rating methods to evaluate \llmc{}s' joint energy-accuracy performance. Specifically, our first rating method, Concentric Incremental Rating Circle (\cic), is a distance-based approach that quantifies inefficiency of each model by the distance of that model to the most optimal achievable objectives. 

We also propose Observation to Expectation Rating (\ote), a parametric model that captures the correlation between energy and accuracy, and evaluate each model's efficiency by comparing observed and expected performance. Building on these methods, 
we assign rating between $1$--$5$ to $22$ models, where the rating $5$ denotes jointly strong accuracy and efficiency and $1$ represents energy-hungry, low-performing models. This five-point scale strikes a balance between sufficient granularity for model differentiation and practical interpretability, consistent with established energy rating frameworks~\cite{aienergyscore-leaderboard,almasri2021star}.

While \framework{} can be broadly applicable across language models, we focus on \llmc{}s as they represent an ideal testbed for this study. 
Unlike natural language, which permits semantic variation while maintaining equivalence, code-related tasks demand strict syntactic and functional correctness along with structured comprehension and analysis---spanning code generation, where outputs must compile and pass tests, and code summarization, where explanations must faithfully reflect program semantics. This constraint enables objective assessment of \llmc{} capabilities.
Moreover, the energy–accuracy characteristics of \llmc{}s remain underexplored, despite their widespread production deployment. This focus equips the SE community with a principled framework for task-specific model selection and fosters the development of greener, more accurate \llmc{}s.

This study makes the following contributions.
\begin{itemize}
    \item We propose two novel rating methods that jointly and systematically use functional correctness and energy efficiency to evaluate LLMs and yield a unified rating.
    \item  We present a comprehensive benchmarking framework for \llmc{}s, \framework{}, dedicated to SE-specific tasks, that simplifies model evaluation and comparison.
    \item The proposed framework and the ranking methods provide a basis for model profiling and lay the groundwork for continued exploration in this area.
    \item We make \framework{} publicly available~\cite{anonBrace2025}. The replication package contains the necessary code and data to replicate, as well as benchmark new models.
\end{itemize}

The remainder of this paper is organized as follows: Section~\ref{sec:methods} presents methodology for the proposed scoring mechanisms, and Section~\ref{sec:results} presents results of the experiments across tasks. Section~\ref{sec:discussion} discusses implications and guidance on the method selection, while Section~\ref{RW} surveys related work. Section~\ref{sec:threats} examines threats to validity, and Section~\ref{sec:conclusion} concludes with future directions.

\section{Methodology}
\label{sec:methods}
\subsection{Problem statement}

Let there be a set of models denoted as
$M = \{m_1, m_2,\dots, m_k\}$,
and a set of SE benchmarks represented as $B = \{b_1, b_2,\dots, b_{k'}\}$. 
Each model is executed on the selected benchmarks. From these runs, we collect the energy consumption profile of each model on three core devices, \ie{} $D = \{cpu,\:gpu,\:ram\}$ along, with its corresponding accuracy score for a given benchmark. We obtain the energy consumption of a model on a benchmark as follows:

\begin{equation}
\begin{gathered}
E_{m,b}\approx\sum_{d\in D}\sum_{n=1}^{N_{m,b}} P_{d,n}^{(m,b)}\,\Delta t, \;\;m\in M,\,b\in B
\end{gathered}
\end{equation}
where $\Delta t$ represents the sampling interval in seconds, $P_{d,n}^{(m,b)}$ shows the power consumption of a device, in watts, and $N_{m,b}$ defines the sampling count.

Subsequently, we define a model rating function $\varphi$ to address the limitations of previous work to unify both energy and accuracy:

\begin{equation}
\begin{gathered}
R_{b} = \varphi_b(A_b,E_b);\;\; s.t. \;\; R_{b,m} \in \{1,..., 5\}
\end{gathered}
\end{equation}
The function $\varphi$ takes the energy-accuracy pairs of all models, \ie{} $A_b$ and $E_b$, from a given benchmark, and outputs a unified energy-accuracy rating for each model on a $1$--$5$. These ratings advocate energy-efficient and highly accurate models from best, \ie{} $5$, to worst, \ie{} $1$, facilitating easier model selection on an SE task.

\begin{figure}[t]
    \centering
    \includegraphics[width=\columnwidth]{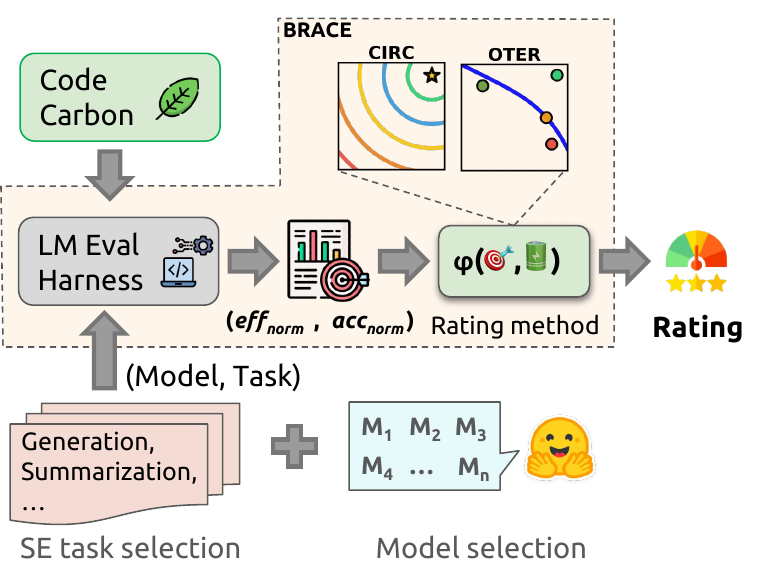}
    \caption{Overview of the \framework{} benchmarking framework. $eff_{norm}$ and $acc_{norm}$ are the normalized values of energy efficiency and accuracy of all models.}
    \label{fig:overview}
\end{figure}

\subsection{Overview}

This study aims to benchmark and rate LLMs based on their combined performance in terms of energy efficiency and accuracy.
We articulate the following research question (RQ) to structure our exploration.

\begin{rqboxgray}
\textbf{RQ:} How can we benchmark LLMs effectively based on their energy-accuracy trade-offs using a unified rating method?  
\end{rqboxgray}
Lasting solutions demand balance; prioritizing one goal in isolation risks undermining accuracy or amplifying environmental impact. Additionally, comparing models solely based on their raw energy–accuracy profiles is challenging, especially when trade-offs exist, \ie{} one model excels in accuracy while consuming significantly more energy, whereas another achieves better efficiency at the expense of precision. Hence, a unified mechanism that integrates both aspects robustly and fairly is essential. 
Fairness ensures that moderately performing models are not penalized, while robustness prevents over-rewarding models that achieve easy gains. 
The RQ explores approaches to unify energy-accuracy trade-offs.

In this study, as outlined in Figure~\ref{fig:overview},
we propose two novel techniques to benchmark LLMs along two dimensions \ie{} energy efficiency and accuracy. 
The accuracy metric is selected based on the chosen task and evaluation criteria.
Our framework, \framework{}, enables us to measure and log both metrics for a given set of models evaluated on a predefined set of tasks and corresponding benchmarking datasets.
The proposed scoring techniques further allow us to rate LLMs on a unified and interpretable scale. 
Rating models on a $1$-$5$ scale enhances interpretability, ensures consistent ranking, and establishes sustainability as the key criterion for ranking models.

\subsection{Benchmark dataset selection}

This study aims to evaluate the capabilities of \llmc{}. Therefore, our primary criterion for selecting benchmarks is their relevance to SE tasks.
Given that the study benchmarks the sustainability characteristics of language models, we select two tasks to analyze model behavior across different problem types.
To evaluate each model’s true capability and resource usage, we focus exclusively on generative tasks and adopt tasks that are widely used in SE research. 

Our first task is \textit{code generation}, which assesses a model's ability to generate code that passes predefined test cases. 
We choose the livecodebench~\cite{jain2025livecodebench} benchmark,
a recent and widely cited dataset that includes diverse code intelligence tasks~\cite{abdollahi2025surveying}.
We use \texttt{pass@1} metric to assess models on this benchmark, which indicates whether the model generates a correct solution on its first attempt~\cite{chen2021evaluating}.
Our second task, \textit{code summarization}, involves generating a docstring for a given code block to assess the model’s code comprehension ability~\cite{sontakke2022code}.
We use the CodeXGLUE benchmark~\cite{lu2021codexglue}, a widely adopted and highly cited dataset~\cite{jelodar2025large,ahmed2022few,afrin2025resource}. 
We adopt its Python code summarization variant and use smoothed blue~\cite{lin2004orange} as the evaluation metric, following the original benchmark setup~\cite{lu2021codexglue}.
Both these tasks complement each other by capturing both text-to-code and code-to-text capabilities of the models.

\subsection{Model selection}

We use HuggingFace~\cite{huggingface}, the largest platform for open-source LLMs, for our model selection search. 
Given the SE focus, we only evaluate \llmc{} in this study.
We only consider open-source \llmc{} so that can be deployed locally to measure their energy consumption accurately.
Our initial search criterion selects models whose names include the term \textit{``code''}, reducing the pool of models from $2,105,378$ to $24,840$. We then filter for \textit{text-generation} models, narrowing the list to $8,840$. 
Our computing environment (\ie{} $32$GB GPU)
can support models up to $9$ billion parameters with full $32$-bit precision.
Therefore, we filter out the models with more than $9$B parameters, resulting in $5,607$ candidates. From the set, we exclude all quantized variants to ensure consistency and fairness and select the top $22$ base models by download count. 
The list of the selected models and their download links can be found in our replication package.

\subsection{Unified scoring mechanisms}
The problem studied in this paper is a classic Multi-Criteria Decision Making (MCDM)~\cite{zionts1979mcdm} with two objectives where we need to choose the optimal result from various alternatives pairs, \ie{} model energy-accuracy pairs.

We reformulate our objective as a standard optimality problem, aiming to maximize both aspects without compromising either. 
To ensure a stable and comparable ratio between them, we operate on min--max normalized values. In this setup, lower energy consumption---indicating higher efficiency---corresponds to a higher normalized value. Hence, we define the \textit{energy efficiency}, denoted as $eff_{norm} = 1 - energy_{norm}$. However, since the accuracy is already positively oriented (\ie{} higher is better), we apply only min-max scaling to normalize it which we denote it as $acc_{norm}$. With this background,
we now define the two novel rating methods for unified energy-accuracy scoring.

\subsubsection*{\textbf{Concentric Incremental Rating Circle (CIRC)}}.

\begin{figure}[h!]
    \centering
    \includegraphics[width=0.9\linewidth]{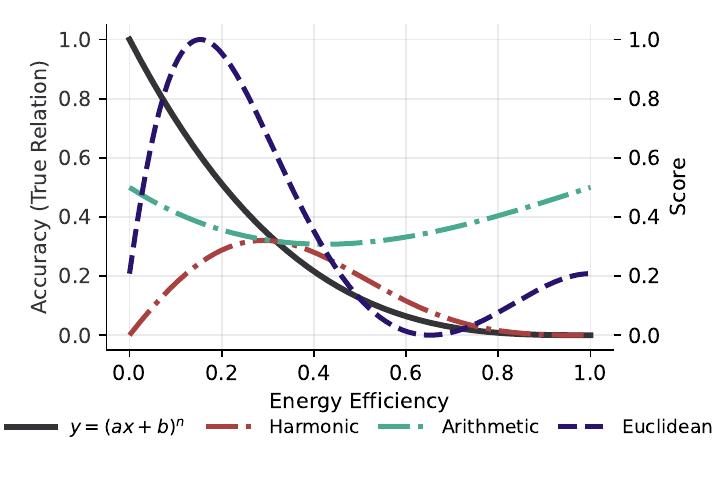}
    \caption{Mean variants and Euclidean distance behavior on a hypothetical relation between energy efficiency and accuracy.}
    \label{fig:mean_limitation}
\end{figure}

As discussed earlier, mean-based variants provide a potential way to jointly account for energy and accuracy. 
Consider a scenario where the accuracy and energy follow a relationship denoted by $y=(ax+b)^n$. 
For this relationship, we plot their mean-based and Euclidean distance-based scores as shown in Figure~\ref{fig:mean_limitation}. 
The arithmetic mean fails to penalize cases with extremely low performance in one objective, \ie{} (0, 1) and (1, 0). Conversely, the harmonic mean penalizes low values too heavily, preventing balanced compensation between the two objectives. In contrast, the Euclidean distance offers a sound trade-off between the arithmetic and harmonic means, enabling moderate and fair compensation. 

Therefore, we propose CIRC, which builds upon Euclidean distance as its core foundation and leverages concepts from TOPSIS~\cite{Hwang1981}--- an MCDM algorithm to rank models by their Euclidean distance to the ideal most optimal and worst optimal solution.

In the first step of CIRC, we define the optimal points for our setting. Unlike TOPSIS, which determines these points relatively, we leverage the normalized nature of our data to fix them at two absolute positions. Specifically, since both objectives are normalized within $0 \le o_1, o_2 \le 1$, we set the best point at $(1, 1)$---representing maximum accuracy and energy efficiency---and the worst point at $(0, 0)$ which are $\sqrt{2}$ away from each other based on Euclidean distance.

To rate models on a $1$-$5$ scale, we partition the interval $[0, \sqrt{2}]$ into five equal segments of length $\frac{\sqrt{2}}{5}$, which serve as score boundaries. Each boundary is defined as a circle to ensure all points on it are at the same distance from the ideal, using the Euclidean-like equation: \((x-c_x)^2+(y-c_y)^2=r^2\).
This setup constructs five concentric circles centered at the ideal point ($1$, $1$), with radii increasing in equal increments, starting from $\frac{\sqrt{2}}{5}$, and ending at $\sqrt{2}$.

After constructing five incremental circles, we assign the highest score to the model that lies within the first circle, \ie{} the circle nearest to the ideal point.
Models with distances between the first and second radii receive a score of four, and this process continues until the region between the fourth and fifth circles, which is assigned a score of one.

\subsubsection*{\textbf{Observation to Expectation Rating (OTER)}}

We propose OTER to address the limitation of linearity assumed by existing unifying approaches, such as arithmetic mean and ratio-based methods. These approaches implicitly assume that an increase in energy consumption should correspond to a proportional, linear increase in accuracy to maintain balance. However, this assumption is fundamentally flawed, as the relationship between energy and accuracy can be nonlinear and vary across models. In fact, improvements in accuracy do not necessarily require energy to increase at the same rate~\cite{mlenergybenchmark-arxiv25,alizadeh2024analyzing}; thus, the true correlation between these two aspects is unknown and must be estimated.

To address previous limitations, we propose OTER method inspired by SFA~\cite{aigner1977formulation}---a parametric method that models input–output relationships to define an optimal frontier---but enhanced with several novel integrations tailored to our problem.

To design the OTER, we first postulate that as models become more energy efficient, a certain degree of accuracy compromise can be acceptable. This assumption offers two key advantages: (1) it allows us to reward models that improve both energy efficiency and accuracy compared to a poorer model, and (2) it enables us to learn the acceptable trade-off rate explicitly, unlike previous approaches that ignore this relationship. To learn this acceptable trade-off, we first reformulate our problem as a 2d problem where $x$ axis shows energy efficiency of each model and $y$ axis is their corresponding accuracy. Then, we fit a monotonically decreasing reference curve that captures the trend observed in the data, \ie{} energy–accuracy pairs---unlike SFA, which estimates the optimal frontier of top performers. In fact, OTER provides an interpretable and robust curve for understanding how energy efficiency and accuracy should co-vary, \ie{} the acceptable trade-off. Also, it informs us at each level of energy efficiency what degree of accuracy we can expect the models to reflect.

Based on this curve, models are rewarded or penalized depending on how their accuracy compares to the expected level, measured as the ratio of actual to expected accuracy. Specifically, models above the expectation curve demonstrate superior performance, as the general trend suggests a lower expected accuracy at their efficiency level, yet they exceed it. In contrast, models below the curve fail to reach the expected accuracy, indicating inefficiency in translating energy consumption into accuracy compared to the dominant behavior observed in other models.
Finally, we rate models on a $1$-$5$ scale by dividing the range between the minimum and maximum of the ratio derived from actual to expected accuracy into five equal intervals and assigning each model to the appropriate interval. 
We formulate this process using the following equation.
\begin{equation}
\label{eq:OTE}
\begin{gathered}
\mathcal{D} = \{(x_i,y_i)\}_{i=1}^n,\\
\operatorname*{arg\,min}_{f\in\mathcal{F}}
  \sum_{i=1}^n \ell\;\!\bigl(y_i, f(x_i)\bigr)
  \quad\text{s.t.}\quad f(x) \; decreasing\ \\
v_i = \frac{y_i}{f(x_i)},\qquad
v_{\min}=\min_{1\le i\le n}v_i,\quad
v_{\max}=\max_{1\le i\le n}v_i,\\[2pt]
\Delta = \frac{v_{\max}-v_{\min}}{5}, \qquad
R_i = \max(1, \lceil \dfrac{v_i-v_{\min}}{\Delta} \rceil)
\end{gathered}
\end{equation}
where $\mathcal{D}$ represents energy ($x$)--accuracy ($y$) pair for all considered models. $f$ represents a monotonically decreasing function that best approximates these data points, by optimizing the objective function of $\ell$. A model's raw score, $v_i$, is defined by dividing the observation, \ie{} $y$, by the expectation, \ie{} $f(x)$. $\Delta$ represents the score interval for each class, and the final rating $R$ is assigned based on which interval a model’s score falls into.

\begin{figure}[ht!]
    \centering
    \includegraphics[width=\linewidth]{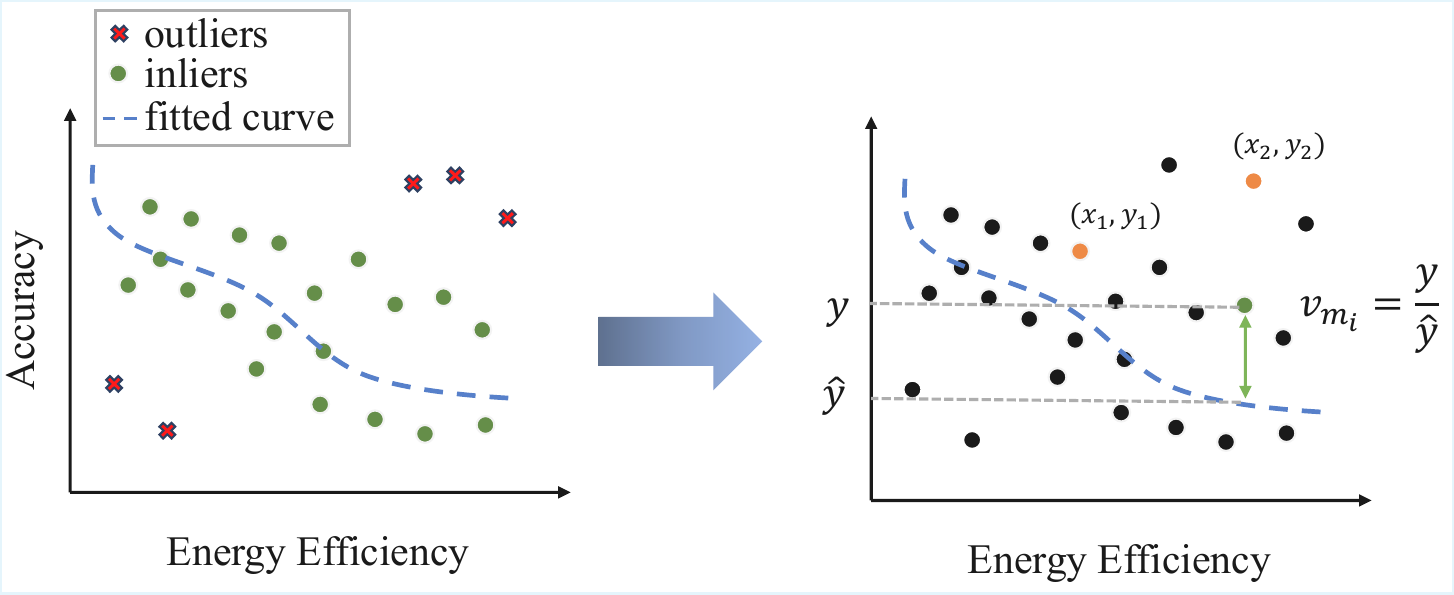}
    \caption{Overview of the OTER approach. We first fit a curve to the majority of the points, followed by computing a raw score $v$ as the ratio of a model’s observed to its expected accuracy.}
    \label{fig:OTE}
\end{figure}

Figure~\ref{fig:OTE} illustrates an overview of the rating approach. As discussed earlier, OTER favors models that maintain or improve accuracy as energy efficiency increases. This property ensures fairness, as a model outperforming another in both objectives should naturally receive a higher score. Such behavior is achieved through the monotonically decreasing nature of the fitted curve. An example of this can be seen in Figure~\ref{fig:OTE} for two points $(x_1, y_1)$ and $(x_2, y_2)$, which can be formally expressed as follows:
\begin{equation}
\begin{gathered}
     f(x) \text{ decreasing},\;\;
V(x,y)=\frac{y}{f(x)},\;\; y_2 \ge y_1 \ge 0.\\
x_2>x_1
\;\Rightarrow\; f(x_2) < f(x_1) \\[2pt]
\;\Rightarrow\; \frac{1}{f(x_2)}> \frac{1}{f(x_1)} \qquad ( f(x)>0)\\[2pt]
\;\Rightarrow\; \,\frac{y_2}{f(x_2)} > \,\frac{y_1}{f(x_1)} 
\;\Rightarrow\; V(x_2,y_2)> V(x_1,y_1).
\end{gathered}
\end{equation}

For models whose accuracy decreases as energy efficiency increases, their raw score 
depends on the curve and their deviation from it.
Points above the curve achieves raw score of more than 1,
while those below score less than 1, indicating poorer performance. Models closer to the curve receive scores closer to 1, placing them near the center of the distribution. Whereas models that deviate most from expectation receive the highest or lowest scores, reflecting exceptionally strong or weak energy-accuracy trade-offs. 
 
According to Equation~\ref{eq:OTE}, two components must be determined to complete the functionality of OTER: first, an optimization algorithm that minimizes the error term $l$ while satisfying the constraint of decreasing monotonicity, and second, the function $f(x)$.

To perform the optimization step, we choose quadratic programming with linear constraints, as it can strictly satisfy the monotonicity constraint in contrast to genetic and machine learning algorithms, and it can also fit any type of smooth function that we desire, unlike Isotonic regression, which yields a step-wise linear~\cite{hardwick2014optimal,Scikit-Isotonic}. Also, it is guaranteed that at least one decreasing solution exists regardless of its proximity to the actual data, enabling quadratic programming to solve the problem where the objective is to minimize the error between predictions and ground-truth data, and linear constraints control the derivations and monotonicity of the curve.

For the function $f(x)$,  we opt for a polynomial function for three main reasons.
First, the problem is bounded on both axes, \ie{} $0 \le x, y\le1$, where polynomials are known to effectively approximate small regions~\cite{bradley2007response}. Additionally, the Stone–Weierstrass theorem ensures that any continuous function on [a, b] can be uniformly approximated by a polynomial~\cite{young2006stone}.
Second, Polynomials can encode monotonicity and convexity in their formulation~\cite{kundu2016call} naturally, making them well-suited for quadratic programming, which requires convex solutions to achieve optimality.
Lastly, polynomial functions are widely used in optimization problems~\cite{kundu2016call,sobrie2018uta,berger2018pareto}. 

Subsequently, since the number of samples in our setting is limited and a few outliers may distort the fitted curve, we remove outliers (as shown in Figure~\ref{fig:OTE}) as an auxiliary step to ensure robustness. We employ Mahalanobis distances computed from a Minimum Covariance Determinant (MCD)~\cite{rousseeuw1984least,rousseeuw1985multivariate} which is highly robust to outliers~\cite{hubert2010minimum} and correlation-aware~\cite{DEMAESSCHALCK20001}. In our settings, observations with squared robust distances exceeding the chi-square threshold at the $95^{th}$ percentile with $2$ degrees of freedom are marked as outliers.

Moreover, we need to determine the variables associated with the monotonic nature of the curve. As we intend to fit the curve in a decreasing way, we must ensure that for all points lying on the curve, their derivatives are always $< 0$. However, when data shows atypical pattern, \ie{} higher energy efficiency corresponds to higher accuracy for majority of the point, the closest fitted function would be a near constant curve. The emergence of such a curve leads to unfair scoring. In fact, if two models show similar accuracy but differ significantly in their energy efficiency, this curve fitting becomes unfair due to the scoring mechanism $y/f(x)$. Since their $y$ values are similar and $f(x)$ is nearly constant, their raw scores will also be similar, even though one model is much more energy efficient and should receive a higher score.

Accordingly to address this issue, \ie{} ensure a fair and robust decreasing slope, we introduce the \textit{Least Expected Slope (LES)}.
Specifically, we compute pairwise slopes for all points that demonstrate an energy-accuracy trade-off, \ie{} demonstrating negative slopes.
Next, we remove extreme derivations and take the third quantile of the remaining values.
Since the slopes are negative, this mechanism offers a conservative upper bound on the expected derivations. We configure the quadratic programming constraints to be less than \textit{LES} in the entire range.
We present \textit{LES} formulation as below:
\begin{equation}
\label{eq:med-def}
\begin{gathered}
    D \;:=\; \Bigl\{\, d_{ij} := \tfrac{y_j - y_i}{x_j - x_i} \;:\; i\neq j,\ x_j\neq x_i,\ d_{ij}<0 \Bigr\}\\
LES \;:=\; Quantile_{0.75}\!\bigl(MCD(D)\bigr),\quad LES < 0
\end{gathered}
\end{equation}

Moreover, we aim to locate the curve always above zero, which can be ensured with a constraint at $f(x=1)$, \ie{} minimum value, due to the monotonicity nature of the function.
Consequently, to fit the most optimal function, satisfy the LES constraint, and ensure the curve remains above zero, we formulate our quadratic programming problem as below:  

\begin{equation}
\label{eq:poly}
\begin{gathered}
\min_{\boldsymbol{\beta}\in\mathbb{R}^{p+1}}
\;\sum_{i=1}^{n}\Bigl(y_i - \sum_{k=0}^{p}\beta_k x_i^{k}\Bigr)^2\\
\text{s.t.}\;\;\sum_{k=1}^{p} k\,\beta_k\,x^{\,k-1} \;\le\; LES
\qquad \forall\,x\in[a,b].\\
\sum_{k=0}^{p} \beta_k\ \;\ge\; \epsilon \quad s.t. \quad \epsilon = 1e^{-2}
\end{gathered}
\end{equation}

In this equation, the first line defines the objective function that minimizes the error between predictions and ground-truth. 
The second line indicates the linear constraints required by quadratic programming. The constraints enforce that each derivative remains less than \textit{LES}. The last constraint ensures the curve always lies above zero. We use a polynomial degree of five to balance the flexibility-stability trade-off: higher degrees may cause instability~\cite{poly_1980}, whereas lower degrees may not capture high variance.

\subsection{Experimental setup}
\label{subsec:setup}
\task{Libraries}
\framework{} leverages two open-source components: Language Model (LM) Evaluation Harness~\cite{eval-harness} and CodeCarbon~\cite{benoit_courty_2024_11171501}. 
The LM Evaluation Harness is widely-used benchmarking platform that supports many applications, such as OpenLLM HuggingFace Leaderboard~\cite{open-llm-leaderboard-v2}.
It provides over $180$ commonly-used benchmarks for LLM evaluation, of which we select LiveCodeBench and CodeXGLUE-Code2Text.
Similarly, CodeCarbon is a well-documented and widely used framework for energy tracking.
It logs the consumed energy asynchronously while the main program runs in the background. 

\task{Machine and Models Configurations}
We set CodeCarbon's sampling interval to one second, aiming to balance sampling frequency and noise during energy consumption measurement~\cite{rajput2024enhancing}. 
Also, we set the tracker mode to \textit{process}, to capture only the consumed energy by the current process.
To ensure measurement reliability, we validated Codecarbon's CPU readings against \textit{perf}, a command-line tool for CPU energy monitoring, and found them consistent.
We also ensured that no other processes were running during the experiments.
To further ensure measurement consistency and account for potential variance, we executed each model-benchmark combination three times.
All models retained their default configurations and hyperparameters, with a fixed random seed for reproducibility.
Each model was offloaded to the GPU during initialization and executed sequentially using the same script. 
Furthermore, no remote pulling and downloading occurred from HuggingFace to maintain consistency.
Lastly, we extracted fine-grained energy measurements across four stages of model execution: model configuration, tokenizer initialization, model initialization, and instance inference.

Our primary machine is equipped with an RTX $5000$ Ada GPU, featuring $32$GB of VRAM, and a $755$ GB AMD EPYC $9554$P $64$-Core processor.
To ensure reproducibility and reliable measurement, we repeated the experiments on a secondary machine featuring an Intel Xeon Gold $5317$ CPU and an RTX $3070$ Ti GPU ($8$ GB VRAM).

\task{Benchmark configurations}
We tested multiple prompts
following prompt engineering best practices~\cite{Anthropicsprompttutorial2025} 
and selected the one yielding the best performance.
For LiveCodeBench, as it is a code generation benchmark, 
generates a complete, test-passing code,
we set the maximum generation length to $1024$ tokens.
For CodeXGLUE, which generates a concise code summary, 
we limited the output to $256$ tokens.
We filtered the datasets to reduce the total runtime to keep the experiments feasible and in reasonable bounds.
Specifically, LiveCodeBench includes over $1000$ samples across two test types, \ie{} standard IO and functional,
and three complexity levels.
We chose $150$ samples; $25$ from each pair of complexity and test cases. We selected first $500$ samples from approximately $14,000$ samples of CodeXGLUE.
With these optimizations,  we executed LiveCodeBench and CodeXGLUE on $22$ models in approximately $13$ and $20$ hours, respectively. 

\task{Framework Extensibility and Usability}
While this study evaluates 22 specific models on two SE tasks, \framework{} is designed for broader applicability beyond our experimental scope. The framework's architecture decouples energy--accuracy evaluation from specific model or benchmark implementations. Users can evaluate any \llmc{} on any benchmark without modifying the framework's code---simply by updating model identifiers and task names in the \framework{} configuration file~\cite{anonBrace2025}. This enables practitioners to benchmark newly released models, evaluate existing models on domain-specific SE tasks, or extend the evaluation to additional code benchmarks with minimal setup overhead.
Moreover, although we adopt the conventional $1$--$5$ rating scale, the framework supports configurable rating ranges (\eg{} $1$--$10$) to accommodate different granularity requirements.

\section{Results}
\label{sec:results}

\begin{table}[t]

\centering

\scriptsize

\setlength{\tabcolsep}{3.5pt}

\renewcommand{\arraystretch}{1.05}
\resizebox{\columnwidth}{!}{
\begin{tabular}{@{}l l |rr|rr|rr|rr@{}}

\toprule

\multirow{2}{*}{\textbf{ID}} & \multirow{2}{*}{\textbf{Model}} & \multicolumn{2}{c|}{\textbf{LCB}} & \multicolumn{2}{c|}{\textbf{CXG}} & \multicolumn{2}{c|}{\textbf{LCB Rate}} & \multicolumn{2}{c}{\textbf{CXG Rate}} \\

\cmidrule(lr){3-4}\cmidrule(lr){5-6}\cmidrule(lr){7-8}\cmidrule(lr){9-10}
& & $A$ & $E$ & $A$ & $E$ & CIRC & OTER & CIRC & OTER \\

\midrule
$\text{M}_{1}$  & deepseek-coder-6.7b-base        & 0.23 & 0.32 & 0.47 & 0.23 
& \cellcolor{green!40} 2 
    & \cellcolor{green!20} 1 
        & \cellcolor{green!80}2
            & 3 \cellcolor{green!60}\\
$\text{M}_{2}$  & starcoderbase-1b                 & 0    & 0.93  & 0    & 0.88 & \cellcolor{green!40} 2 & \cellcolor{green!20} 1 & \cellcolor{green!40} 2 & \cellcolor{green!20} 1 \\
$\text{M}_{3}$  & starcoder2-3b                   & 0.02 & 0.23 & 0.37 & 0.32 & \cellcolor{green!20} 1 & \cellcolor{green!20} 1 & \cellcolor{green!40} 2 & \cellcolor{green!60} 3 \\
$\text{M}_{4}$  & CodeLlama-7b-Instruct-hf        & 0.21 & 0.82 & 0.76 & 0.23 & \cellcolor{green!60} 3 & \cellcolor{green!20} 1 & \cellcolor{green!60} 3 & \cellcolor{green!80} 4 \\
$\text{M}_{5}$  & Qwen2.5-Coder-3B-Instruct        & 0.52 & 0.96  & 0.36 & 0.62 & \cellcolor{green!80} 4 & \cellcolor{green!60} 3 & \cellcolor{green!60} 3 & \cellcolor{green!60} 3 \\
$\text{M}_{6}$  & Qwen2.5-Coder-7B-Instruct       & 0.38 & 0.78 & 0.36 & 0.76 & \cellcolor{green!60} 3 & \cellcolor{green!40} 2 & \cellcolor{green!60} 3 & \cellcolor{green!60} 3 \\
$\text{M}_{7}$  & Qwen2.5-Coder-1.5B               & 0.48 & 1  & 0.31 & 0.90 & \cellcolor{green!80} 4 & \cellcolor{green!60} 3 & \cellcolor{green!60} 3 & \cellcolor{green!60} 3 \\
$\text{M}_{8}$  & deepseek-coder-1.3b-base         & 0.10 & 0.78 & 0.70 & 0.81 & \cellcolor{green!40} 2 & \cellcolor{green!20} 1 & \cellcolor{green!80} 4 & \cellcolor{green} 5 \\
$\text{M}_{9}$  & deepseek-coder-7b-instruct & 0.58 & 0.77 & 0.35 & 0.35 & \cellcolor{green!80} 4 & \cellcolor{green!60} 3 & \cellcolor{green!40} 2 & \cellcolor{green!40} 2 \\
$\text{M}_{10}$ & starcoder2-7b                   & 0.04 & 0.15 & 0.56 & 0.21 & \cellcolor{green!20} 1 & \cellcolor{green!20} 1 & \cellcolor{green!40} 2 & \cellcolor{green!60} 3 \\
$\text{M}_{11}$ & codegen-350M-mono               & 0.02 & 0.86  & 0.09 & 0.86  & \cellcolor{green!40} 2 & \cellcolor{green!20} 1 & \cellcolor{green!40} 2 & \cellcolor{green!20} 1 \\
$\text{M}_{12}$ & Qwen2.5-Coder-0.5B              & 0.27 & 0.99   & 0.37 & 0.91  & \cellcolor{green!60} 3 & \cellcolor{green!40} 2 & \cellcolor{green!60} 3 & \cellcolor{green!60} 3 \\
$\text{M}_{13}$ & Yi-Coder-9B                     & 0.33 & 0    & 0.16 & 0    & \cellcolor{green!20} 1 & \cellcolor{green!20} 1 & \cellcolor{green!20} 1 & \cellcolor{green!20} 1 \\
$\text{M}_{14}$ & Replete-Coder-Llama3-8B & 0.21 & 0.74 & 0.20 & 0.28 & \cellcolor{green!60} 3 & \cellcolor{green!20} 1 & \cellcolor{green!40} 2 & \cellcolor{green!40} 2 \\
$\text{M}_{15}$ & speechless-code-mistral-7b & 0.31 & 0.85 & 0.37 & 0.60 & \cellcolor{green!60} 3 & \cellcolor{green!40} 2 & \cellcolor{green!60} 3 & \cellcolor{green!60} 3 \\
$\text{M}_{16}$ & stable-code-3b                   & 0.06 & 0.68 & 0.53 & 0.68 & \cellcolor{green!40} 2 & \cellcolor{green!20} 1 & \cellcolor{green!60} 3 & \cellcolor{green!80} 4 \\
$\text{M}_{17}$ & CodeLlama-7b-Python-hf          & 0.15 & 0.73 & 0.38 & 0.23 & \cellcolor{green!40} 2 & \cellcolor{green!20} 1 & \cellcolor{green!40} 2 & \cellcolor{green!40} 2 \\
$\text{M}_{18}$ & Seed-Coder-8B-Instruct          & 1 & 0.88  & 0.31 & 0.59 & \cellcolor{green} 5 & \cellcolor{green} 5 & \cellcolor{green!60} 3 & \cellcolor{green!40} 2 \\
$\text{M}_{19}$ & CodeQwen1.5-7B-Chat             & 0.06 & 0.99  & 0.03 & 1 & \cellcolor{green!40} 2 & \cellcolor{green!20} 1 & \cellcolor{green!40} 2 & \cellcolor{green!20} 1 \\
$\text{M}_{20}$ & Magicoder-S-DS-6.7B             & 0.42 & 0.61 & 0.39 & 0.42 & \cellcolor{green!60} 3 & \cellcolor{green!40} 2 & \cellcolor{green!60} 3 & \cellcolor{green!60} 3 \\
$\text{M}_{21}$ & granite-8b-code-base-4k         & 0.10 & 1  & 1 & 0.14 & \cellcolor{green!40} 2 & \cellcolor{green!20} 1 & \cellcolor{green!40} 2 & \cellcolor{green} 5 \\
$\text{M}_{22}$ & codegen-2B-mono                  & 0.02 & 0.68 & 0.45 & 0.57 & \cellcolor{green!40} 2 & \cellcolor{green!20} 1 & \cellcolor{green!60} 3 & \cellcolor{green!60} 3 \\

\bottomrule

\end{tabular}
}

\caption{
Summary of model performance on LiveCodeBench (LCB) and CodeXGLUE (CXG): normalized accuracy ($A$), normalized efficiency ($E$) and unified ratings (CIRC and OTER) for each benchmark.
}

\label{tab:results}
\end{table}

In this section, we present the evaluation results of all models and compare their performance on the unified scale of energy-accuracy using our rating methods.

\task{Comparative Analysis on Two Benchmarks}
Table~\ref{tab:results} presents accuracy and energy efficiency reported for all the considered models for both the tasks.
The table also shows the rating assigned by CIRC and OTER for all the models, along with their IDs.
Seed-Coder-8B-Instruct ($M_{18}$),
achieves a perfect score on LiveCodeBench and a moderate one on CodeXGLUE, making it the most optimal overall. This observation reveals that this model, built on the Llama backbone, efficiently converts its computational capacity into accuracy.  
Following it are the Qwen models ($M_5$ and $M_7$),
both attaining moderately high ratings on both LiveCodeBench and CodeXGLUE tasks. In fact, these are the only models that do not sacrifice one objective completely over the other, balancing a sound trade-off between them.
Interestingly, in contrast, the largest Qwen model ($M_6$) performs poorly than the smaller size variants on LiveCodeBench,
highlighting that larger models does not guarantee higher accuracy. 
On the other hand, the weakest model in both tasks is Yi-Coder-9B ($M_{13}$), which shows the lowest energy efficiency and moderate accuracy, indicating inefficient parameter utilization.
To statistically validate that ratings reflect genuine energy--accuracy trade-offs rather than systematic size bias, we conduct Kruskal-Wallis tests---a non-parametric test suitable for comparing rating distributions across groups without assuming normality. We group models by parameter size 
($<3$B: n=5, 3--7B: n=6, $\ge7$B: n=11) and test the null hypothesis that all size groups exhibit the same rating distribution. Results across all rating method--benchmark combinations fail to reject the null hypothesis ($p>0.05$ for all: LCB-CIRC $p=0.84$, LCB-OTER $p=0.97$, CXG-CIRC $p=0.17$, CXG-OTER $p=0.62$), confirming no significant size-based bias.

\begin{figure}[t] 
  \centering
  \includegraphics[width=\linewidth]{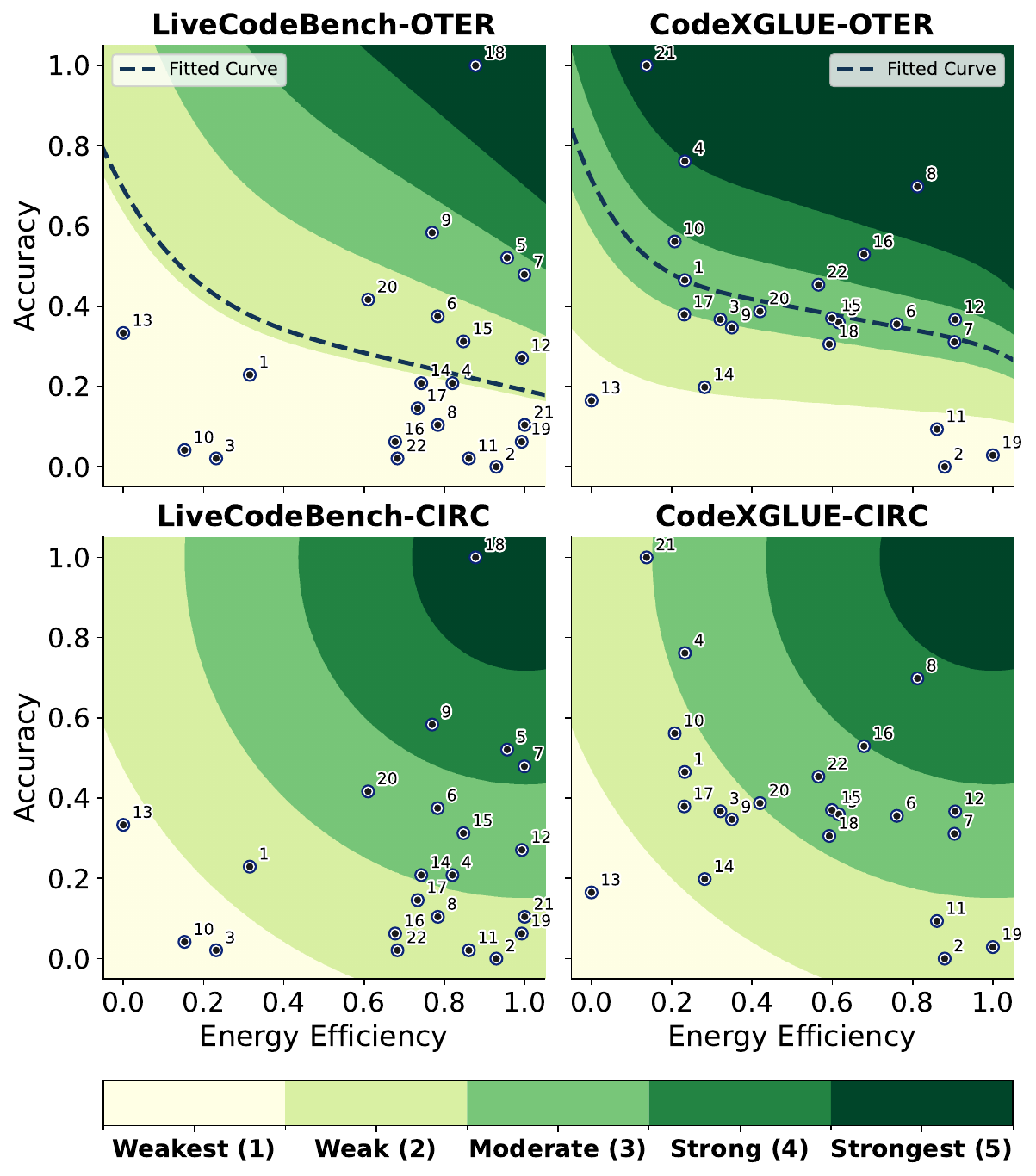}
  \caption{Unified energy-accuracy model rating; numbers in the plot show models' IDs}
  \label{fig:RQ1_main_plots}
\end{figure}

\task{Rating Methods Behavior Interpretation}
We interpret how each rating method rates models based on their observed performance, shown in Figure ~\ref{fig:RQ1_main_plots}.
In the top-left plot, we observe OTER fits a curve to estimate the trend. It ranks over half the models in the weakest class.
This results from the highly capable model $M_{18}$, which achieves perfect accuracy and near-optimal energy efficiency. Despite the presence of outliers and models exhibiting high energy efficiency but low accuracy, OTER does not favor them, demonstrating robustness.

The right-top plot demonstrates OTER rates the majority of models to middle classes as they reflect a balanced trade-off between energy efficiency and accuracy.
The representation shows robustness at the highest energy efficiency, where moderate accuracy is expected, unlike $ M_{2}$, $M_{11}$, and $M_{19}$.
It also fairly rewards the most accurate model with relatively low energy efficiency ($M_{21}$), assigning it the highest score as it is considered as an exceptional outlier.

To statistically validate OTER's robustness to design choices, we conduct systematic sensitivity analyses on two critical aspects: hyperparameter selection and outlier removal. OTER's methodology involves explicit design choices—polynomial degree (5), MCD outlier threshold (95th percentile), LES quantile (75th), and $\epsilon$ (0.01) at f(x=1)---that could potentially introduce fragility. We test 358 hyperparameter configurations varying polynomial degree $\in\{3,4,5,6,7\}$, MCD percentile $\in\{90, 95, 97.5\}$, LES quantile $\in\{65, 70, 75, 80\}$, and $\epsilon$ $\in \{{0.001, 0.01, 0.1}\}$ computing Spearman correlations between each configuration's ratings and our baseline. Results demonstrate exceptional stability: mean $\rho=0.985$ (min $\rho=0.922$), with all 358 configurations achieving $\rho>0.85$ and maintaining 96.5\% average rating stability (min 81.8\%). All correlations are statistically significant ($p<0.001$). To assess the impact of outlier removal, we compare ratings with and without MCD-based outlier filtering. Both configurations maintain a strong correlation ($\rho>0.959$) with 88.6\% average stability, confirming that outlier removal enhances robustness without introducing systematic bias. These results validate that OTER's ratings reflect genuine energy--accuracy patterns rather than artifacts of specific hyperparameter choices, confirming the reliability of the methodology across reasonable parameter variations.
 
The bottom left plot demonstrates the CIRC approach, which rates models regarding the circle that they fall into. Since this is a Euclidean-based ranking, both objectives are considered equally. 
This plot is predominantly dense in the weak class, as most models perform well on only one objective. The absolute distances along each curve allow fair comparison among models within the class.
The bottom right plot also exhibits similar behavior. As a static scoring approach, CIRC leaves the very strong class empty, highlighting room for model improvement.

Having validated OTER's hyperparameter robustness and confirmed size-agnostic fairness for both methods, we further assess rating stability and fairness under two practical deployment scenarios: whether individual models disproportionately influence the overall ratings, and whether measurement noise in energy--accuracy inputs destabilizes rankings. We conduct two targeted tests addressing these concerns.
First, we perform leave-one-out (LOO) analysis, removing each model sequentially and recomputing ratings for the remaining models to assess dependency on individual data points. Results demonstrate high stability for both methods: CIRC achieves a mean rating drift of 0.024 (LCB) and 0.022 (CXG) with Kendall-$\tau > 0.98$ for both tasks, while OTER maintains a mean drift of 0.035 (LCB) and 0.071 (CXG) with Kendall-$\tau > 0.95$, confirming that ratings remain consistent regardless of which model is excluded. Second, we test measurement noise robustness by perturbing energy-accuracy values with uniform noise ($\epsilon = \pm 5\%$) across 20 trials. Both methods show resilience: CIRC exhibits mean rating changes of 0.014 (LCB) and 0.050 (CXG), while OTER shows 0.041 (LCB) and 0.111 (CXG), with worst-case changes of 1 rating class. These results confirm that both CIRC and OTER produce stable, reliable ratings robust to individual model influence and measurement uncertainty.

\task{Difference in OTER and CIRC on LiveCodeBench}
In this section, we compare how OTER behaves differently from CIRC. 
We can observe in Table ~\ref{tab:results} and the
first column of 
Figure~\ref{fig:RQ1_main_plots} that the order of scores in both approaches are similar, with most models differing by only one class. In fact, CIRC is more lenient in scoring than OTER, assigning higher ranks to models that demonstrate at least one high objective. 
Nonetheless, we can conclude that both methods rank fairly with equal orders, only varying in the middle classes. It is also worth mentioning that due to the discretizing effect, we can observe many models that are close to boundaries, which a minor change in one of the values would cause their class to change. This fact can also lead to a slight variance in both of our proposed methods.

\task{Difference in OTER and CIRC on CodeXGLUE}
For the CodeXGLUE benchmark, we observe more conflicts by analyzing the CXG rate in Table~\ref{tab:results}.
This behavior arises from the relations that are captured by OTER. 
For instance, analyzing the right column of Figure~\ref{fig:RQ1_main_plots} granite-8b-code-base-4k, \ie{} $M_{21}$, is the model that is rated as ``Strongest'' in OTER, emphasize the severity of achieving high accuracy, while it is ranked as ``Weak'' in the CIRC approach, indicating its low energy efficiency. 
Furthermore, similar to OTER in LiveCodeBench, OTER in CodeXGLUE does not favor the highest energy-efficient models. 
This occurs because most models cluster within moderate accuracy ($0.3$--$0.6$) while spanning the entire energy efficiency spectrum, revealing that high efficiency is easier to achieve than high accuracy, as few models attain high accuracy regardless of their efficiency level. Overall, both methods are performing closely to each other
with some exceptions that highlight their distinct behaviors depending on the perspective taken.

\task{Benchmark Effects on Ratings}

In the first row of Figure~\ref{fig:RQ1_main_plots}, showing OTER results,
scores are generally lower on LiveCodeBench. This reflects the higher complexity of generating structured, grammar-based text that also passes unit tests, compared to open-ended text generation with a less rigid evaluation metric. 
Our novel ranking method captures this subtle point, with OTER remains trend-normalized per task and assigning higher expected accuracy to CodeXGLUE without trivially favoring the easier task. 
The second row in the Figure~\ref{fig:RQ1_main_plots} shows model ratings using CIRC.
While models on CodeXGLUE have higher average accuracy, reflecting the task's relative ease, their low energy efficiency places most models
in the second and third classes.

Moreover, variations in model ratings can be analyzed by comparing their scores under the LCB and CXG rates in Table~\ref{tab:results}.
In general, variances are more pronounced in the OTER approach, where the fitted curve produces model-specific scoring. We highlight a few largest differences. 
The first model demonstrating the largest variance, deepseek-coder-1.3b-base ($M_8$), performs well on CodeXGLUE, achieving $0.7$ accuracy and $0.81$ energy efficiency, leading to scores of $4$ and $5$.
However, on LiveCodeBench, its accuracy drops to $0.1$ despite a reasonable energy efficiency of $0.78$, achieving scores of $2$ (CIRC) and $1$ (OTER).
This behavior is expected because OTER interprets high energy efficiency is relatively easier to achieve than high accuracy. 
Therefore, being a relatively small, the model struggles on the more demanding LiveCodeBench and leads to a large difference in scores across the two tasks. 
Another example of large difference in the scores is Seed-Coder-8B-Instruct ($M_{18}$). This model shows high energy efficiency on both LiveCodeBench and CodeXGLUE. However, the primary cause of its score difference is accuracy. It scores a perfect $5$ on LiveCodeBench, but only $3$ (CIRC) and $2$ (OTER) on CodeXGLUE.
This suggests that the model excels at code generation and completion but is less capable in code comprehension tasks. 

\begin{rqbox}[Summary]
\begin{itemize}[leftmargin=*, labelsep=4pt]
    \item Models generally scored better in CodeXGLUE because of the easy nature of the task
    \item Since model accuracy shows low variance across most data points, OTER interprets high accuracy as harder to achieve and therefore rewards it more.
    \item CIRC showed as a reliable and deterministic algorithm, while OTER showed as a trend-aware and sensitive to outliers, which makes it robust.
    \item  Trade-offs in CIRC are static while trade-off in OTER depends on the position of the point
    \item Models final scores can largely vary between two tasks in some cases, which is dependent on architectures and the pretraining phase, specifically how much emphasis was placed on code comprehension versus generation.
\end{itemize}
\end{rqbox}

\section{Discussions}
\label{sec:discussion}

Based on the characteristics of CIRC and OTER, 
both approaches can effectively solve the MCDM problem,
though their practicality depends on the users’ criteria.

CIRC allows compensation between objectives without heavily penalizing low performance in one, favoring both aspects equally. Hence, it is an ideal choice when we cannot estimate the correlation of objectives.
Additionally, it provides robust scoring by fixing an optimal point, ranking models based on their distance from it. 
The pre-defined class ranges ensure that a model’s score is consistent regardless of other samples’ performance. 
This method remains consistent when new models are added.
Specifically, if the new data's accuracy or energy efficiency falls within existing ranges, scoring remains unchanged.
However, if it exceeds current limits,  the scores must be recalculated to ensure fair scoring.

OTER, on the other hand, is a trend-aware mechanism that rewards models performing better than expected.
Hence, depending on the expectation curve,
it can reward models excelling in one objective, demonstrating fairness, and penalize those with moderate performance in both objectives, demonstrating robustness. 
A key advantage of OTER is its ability to reveal each model’s true capacity, highlighting outliers and guiding further improvement, as reflected in the unbounded raw scores, \ie{} $0 \le v \le \infty$. 
Similar to CIRC, OTER provides a fair score when new data points are added to our dataset. 
However, since it is a trend-aware and pre-normalized method, it is recommended to refit the model to ensure consistent comparisons across models.

Lastly, although we demonstrated two criteria decision-making in this work, the approach can be generalized to an arbitrary number of objectives. In essence, all objectives must first be aligned in the same maximization direction. Subsequently, we need to create a two-objective MCDM based on each pair of objectives. 
Finally, we require to derive a score for each pair of objectives and average them across all pairs.

\section{Implications}
\framework{} empowers users of LLMs and AI technologies, in our context software development teams,
to take actionable steps toward improving their workflows and adopting AI solutions that are both accurate and energy-efficient.
\textit{Model hosting platforms} such as HuggingFace can integrate the proposed energy–accuracy ratings alongside their listed models to promote more informed and sustainable adoption choices. \textit{Software developers} and \textit{GenAI practitioners} can confidently select the most optimal models for their projects, balancing performance with environmental impact. Similarly, companies can leverage these ratings during product development to position themselves as both performance-driven and environmentally conscious, thereby attracting customers. \textit{Researchers} can also use the framework to analyze architectural differences, identify factors influencing energy–accuracy trade-offs, and uncover gaps across models for future improvement. Although our implementation focuses on \llmc{}, the framework can be readily extended to domain-specific models, enabling fairer comparisons within other fields. Furthermore, it facilitates model benchmarking for \textit{publishers}, helping them understand where their model stands in the broader landscape.

\section{Related Work}
\label{RW}

\subsection{General LLM Benchmarking}
Numerous studies benchmark LLMs based on various criterias\cite{srivastava2023beyond,zheng2023judging}. For example, IFEVAL ~\cite{zhou2023instructionfollowingevaluationlargelanguage} evaluates models on instruction-following abilities using $25$ verifiable tasks.
Similarly, HELM ~\cite{liang2022holistic} provides a holistic bench measuring accuracy, calibration, robustness, fairness, bias, toxicity, and efficiency for $30$ models.  

\subsection{Sustainability Behavior Analysis}
In parallel to evaluating LLMs on different benchmarks, studying the sustainability behavior of current LLMs and optimizing them have been widely investigated in recent literature~\cite{luccioni2023estimating,McDonald_2022,samsi2023words}.
Morrison \etal{}~\cite{morrison2025holistically} holistically measured the resource exhausted throughout the development phase. Moreover, Google and Meta disclose their carbon footprint during their pretraining stages for the Gemma and Llama models, respectively ~\cite{team2403gemma,touvron2023llama}.
Other studies~\cite{rajput2025tu, saad2025adaptive} have examined the impact of optimization techniques on the energy consumption of AI pipeline stages, ranging from pre-training to inference phases.

\subsection{LLM Sustainability and Benchmarking}
Several studies have focused on benchmarking LLMs on their environment impact.
Argerich \etal{} ~\cite{argerich2024measuring} developed an energy profiler  to measure energy consumption of an LLM during inference.
Rajput \etal{} ~\cite{Rajput2024} benchmarked quantization techniques from an energy-efficiency perspective, providing insights for choosing sustainable models in resource-constrained settings.
Samsi \etal{} ~\cite{touvron2023llama} analyzed Llama variants ($7$B, $13$B, $65$B) across different hardware and configurations to highlight the benefits of power capping and improved GPU utilization.
Ecologits ~\cite{EcoLogitsDocs_latest} proposed a framework to estimate the environmental impact of proprietary models, such as GPT-4 or Claude-Haiku, using regression-based energy and latency predictions. 
HuggingFace AI energy score benchmarked both proprietary and open-source models on ten tasks and reported $1$-$5$ star ratings for their energy performance ~\cite{aienergyscore-leaderboard}.

\subsection{Joint Accuracy and Energy Benchmarking}
As already discussed, benchmarking LLMs only on green AI perspective overlooks the functional correctness, making their adoption challenging since their true performance remains unexplored. 
We observe a few works targeting both accuracy and energy aspects. Luccioni \etal{}~\cite{luccioni2024power}  evaluated whether a model exhibits comparable performance regarding its emissions. 
ZEUS~\cite{zeus-nsdi23} experimented with various language models and tasks and formulated a leaderboard with energy-accuracy related metrics. 
Other works study Pareto-optimal energy-accuracy trade-offs in specialized domains: Alizadeh \etal{}~\cite{alizadeh2024analyzing} focus on SE tasks, while Husom \etal{}~\cite{husom2025sustainable} target embedded systems.
Lastly, in the domain of scoring and ranking LLMs from a sustainability perspective, Pham \etal{} ~\cite{pham2025slm} scrutinize sustainability, correctness, and computation characteristics of Small Language Models (SLM). 
Kaplan \etal{} ~\cite{kaplan2025} 
proposed a novel metric, \ie{} delta carbon emission by computing the ratio of normalized carbon emission and f1-score. 
Jegham \etal{} ~\cite{jegham2025hungry} proposed a framework to predict sustainability data of proprietary models.

Despite these advancements in the green AI field, a sound approach that jointly considers both energy and accuracy while capturing their correlated behavior remains lacking. Moreover, the sustainability aspects of \llmc{}s remain underexplored, limiting our ability to systematically compare their energy-accuracy trade-offs.  
Our framework addresses these gaps by providing seamless energy and accuracy computation on two widely used software engineering benchmarks. We introduced two tailored rating methods that classify models on a $1$--$5$ scale based on a fair and robust assessment of both energy and accuracy that can be applied to diverse use cases. We then benchmarked all \llmc{}s using these approaches to reveal the strengths and weaknesses of each model.

\section{Threats to Validity}
\label{sec:threats}

\textbf{Internal Validity} Several factors within our experimental design could introduce measurement variations. For energy measurement reliability, we followed best practices for energy consumption measurement~\cite{rajput2024enhancing}, executed each experiment three times, 
and configured CodeCarbon's tracking mode to \textit{process} to isolate our running process, and cross-validated with the \textit{perf} CLI tool. We measure energy consumption only for code blocks that directly involve LLM operations, excluding any overhead costs.
To prevent GPU and CUDA cache residues from affecting measurements, we executed each model in a separate process rather than within the same process.

\textbf{External Validity} Several factors, such as different subset selection from the benchmarking dataset and hardware differences, may limit generalizability to other contexts. 
However, we selected a reasonably representative subset, providing the criteria used for the selection to reduce the threat.
We note that model outputs can differ
across machines with identical seeds and deterministic CUDA algorithms due to the $16$-bit precision models. In fact, performing the computations in lower precision inherently involve small precision losses, and different machines or CUDA versions handle these operations differently. However, to ensure consistent and fair experimental conditions across all models, we maintained the original precision settings.

\textbf{Construct Validity} 
Both proposed approaches normalize energy and accuracy using the minimum and maximum values across the evaluated models. 
In this process, extreme values can distort relative positions.
However, we introduced reliability measures such as outlier removal. The relatively large set of evaluated models, along with their consistent ratings and analyses, supports the reliability of both approaches.

\section{Conclusions and Future Work}
\label{sec:conclusion}

We propose \framework{} to benchmark \llmc{}s based on their demonstrated functional correctness and energy efficiency.
We also introduce two novel systematic rating methods, CIRC and OTER, to rate \llmc{}s on unified dimensions of accuracy and energy efficiency.
We experimented with
two software engineering benchmarks namely, LiveCodeBench and CodeXGLUE,
to investigate their behavior on two complementary tasks of code-to-text and text-to-code. 
Our experiments and analysis reveal that CIRC serves as a reliable and deterministic method, while OTER is a trend-aware, robust, and sensitive to outliers.
The ratings from this study can serve as an energy–accuracy tag across platforms, helping software and AI practitioners identify models that are both highly accurate and energy-efficient.

We plan to extend this study by investigating the key factors that  \textit{strongly influence model accuracy}, offering valuable insights for future model developers to prioritize in their training stages. 
Also, we intend to delve into the \textit{scaling laws} in terms of energy and accuracy, and whether models are efficiently utilizing their parameters to transform them into performance.
Additionally, we aim to adopt more tasks to generate a single score for each model based on the aggregation of scores. Furthermore, for OTER, we desire to use more models to generate a universal task-specific function that can help the providers target energy efficiency based on their accuracy or vice versa. Lastly, we aim to provide an informative list of optimization techniques to obtain a better rating by investigating the effect of different factors, including hyperparameters and quantization.

\section{Data Availability}

The implemented software is made public, along with instructions on how to run it and our obtained results, in our replication package~\cite{anonBrace2025}.

\bibliographystyle{IEEEtran} 
\balance
\bibliography{references}

\end{document}